 %
 %
\input harvmac.tex  
%
%
 %
\catcode`@=11
\def\rlx{\relax\leavevmode}                   
 %
 %
 %
\font\eightrm=cmr8 \font\eighti=cmmi8 \font\eightsy=cmsy8
  
\skewchar\eighti='177 \skewchar\eightsy='60
%

 %
\font\tenmib=cmmib10
\font\sevenmib=cmmib10 at 7pt 
\font\fivemib=cmmib10 at 5pt  
\font\tenbsy=cmbsy10
\font\sevenbsy=cmbsy10 at 7pt 
\font\fivebsy=cmbsy10 at 5pt  
\def\BMfont{\textfont0\tenbf \scriptfont0\sevenbf
                              \scriptscriptfont0\fivebf
            \textfont1\tenmib \scriptfont1\sevenmib
                               \scriptscriptfont1\fivemib
            \textfont2\tenbsy \scriptfont2\sevenbsy
                               \scriptscriptfont2\fivebsy}
\def\BM#1{\rlx\ifmmode\mathchoice
                      {\hbox{$\BMfont#1$}}
                      {\hbox{$\BMfont#1$}}
                      {\hbox{$\scriptstyle\BMfont#1$}}
                      {\hbox{$\scriptscriptstyle\BMfont#1$}}
                 \else{$\BMfont#1$}\fi}
 %
 %
 %
 %
\def\inbar{\vrule height1.5ex width.4pt depth0pt}
\def\sinbar{\vrule height1ex width.35pt depth0pt}
\def\ssinbar{\vrule height.7ex width.3pt depth0pt}
\font\cmss=cmss10
\font\cmsss=cmss10 at 7pt
\def\ZZ{\rlx\leavevmode
             \ifmmode\mathchoice
                    {\hbox{\cmss Z\kern-.4em Z}}
                    {\hbox{\cmss Z\kern-.4em Z}}
                    {\lower.9pt\hbox{\cmsss Z\kern-.36em Z}}
                    {\lower1.2pt\hbox{\cmsss Z\kern-.36em Z}}
               \else{\cmss Z\kern-.4em Z}\fi}
\def\Ik{\rlx{\rm I\kern-.18em k}}  
\def\IC{\rlx\leavevmode
             \ifmmode\mathchoice
                    {\hbox{\kern.33em\inbar\kern-.3em{\rm C}}}
                    {\hbox{\kern.33em\inbar\kern-.3em{\rm C}}}
                    {\hbox{\kern.28em\sinbar\kern-.25em{\sevenrm C}}}
                    {\hbox{\kern.25em\ssinbar\kern-.22em{\fiverm C}}}
             \else{\hbox{\kern.3em\inbar\kern-.3em{\rm C}}}\fi}
\def\IP{\rlx{\rm I\kern-.18em P}}
\def\IR{\rlx{\rm I\kern-.18em R}}
\def\Ione{\rlx{\rm 1\kern-2.7pt l}}
 %
 %

 %

\def\intem#1{\par\leavevmode%
              \llap{\hbox to\parindent{\hss{#1}\hfill~}}\ignorespaces}
 %


 %
\newskip\humongous \humongous=0pt plus 1000pt minus 1000pt   
\def\caja{\mathsurround=0pt}
\newif\ifdtup
 %
\def\cmath#1{\,\vcenter{\openup2\jot \caja
     \ialign{\strut \hfil$\displaystyle{##}$\hfil\crcr#1\crcr}}\,}
 %
\def\eqalign#1{\,\vcenter{\openup2\jot \caja
     \ialign{\strut \hfil$\displaystyle{##}$&$
      \displaystyle{{}##}$\hfil\crcr#1\crcr}}\,}
 %

 %
\def\panorama{\global\dtuptrue \openup2\jot \caja
     \everycr{\noalign{\ifdtup \global\dtupfalse
      \vskip-\lineskiplimit \vskip\normallineskiplimit
      \else \penalty\interdisplaylinepenalty \fi}}}
 %
\def\cmathno#1{\panorama \tabskip=\humongous
     \halign to\displaywidth{\hfil$\displaystyle{##}$\hfil
       \tabskip=\humongous&\llap{$##$}\tabskip=0pt\crcr#1\crcr}}
 %
\def\eqalignno#1{\panorama \tabskip=\humongous
     \halign to\displaywidth{\hfil$\displaystyle{##}$
      \tabskip=0pt&$\displaystyle{{}##}$\hfil
       \tabskip=\humongous&\llap{$##$}\tabskip=0pt\crcr#1\crcr}}
 %

 %
\def\twoeqsalignno#1{\panorama \tabskip=\humongous
     \halign to\displaywidth{\hfil$\displaystyle{##}$
      \tabskip=0pt&$\displaystyle{{}##}$\hfil
       \tabskip=0pt&\hfil$\displaystyle{##}$
        \tabskip=0pt&$\displaystyle{{}##}$\hfil
         \tabskip=\humongous&\llap{$##$}\tabskip=0pt\crcr#1\crcr}}
 %

 %
 %
 %
 %
\def\,{\hskip1.5pt}           
 %

\let\d=\delta       \let\vd=\partial             
     
\let\f=\phi                       
                                    \let\G=\Gamma
\let\h=\eta

\let\j=\psi                                      
\let\k=\kappa
                                   
\let\m=\mu
\let\n=\nu

\let\r=\rho         
\let\s=\sigma                   

\let\w=\omega

 %
 %
\def\Box{{\sqcap\mkern-12mu\sqcup}}
\def\lapp{\lower.4ex\hbox{\rlap{$\sim$}} \raise.4ex\hbox{$<$}}
\def\gapp{\lower.4ex\hbox{\rlap{$\sim$}} \raise.4ex\hbox{$>$}}
\def\con{\ifmmode\raise.1ex\hbox{\bf*}
          \else\raise.1ex\hbox{\bf*}\fi}
\let\iff=\leftrightarrow

\def\dual{\relax\leavevmode\lower.9ex\hbox{\titlerms*}}
\def\define{\buildrel\rm def\over =}
\let\id=\equiv
\let\8=\otimes
 %
 %
 %
 %
\let\ba=\overline
\let\2=\underline

 %
\def\dt#1{{\buildrel{\smash{\lower1pt\hbox{.}}}\over{#1}}}

\font\eightrm=cmr8
\def\6(#1){\relax\leavevmode\hbox{\eightrm(}#1\hbox{\eightrm)}}
\def\0#1{\relax\ifmmode\mathaccent"7017{#1}     
                \else\accent23#1\relax\fi}      
\def\7#1#2{{\mathop{\null#2}\limits^{#1}}}      
\def\5#1#2{{\mathop{\null#2}\limits_{#1}}}      
 %

\def\ket#1{\left| #1\right\rangle}

 %

 %

 %

 %
\newbox\t@b@x
\def\rightarrowfill{$\m@th \mathord- \mkern-6mu
     \cleaders\hbox{$\mkern-2mu \mathord- \mkern-2mu$}\hfill
      \mkern-6mu \mathord\rightarrow$}
\def\tooo#1{\setbox\t@b@x=\hbox{$\scriptstyle#1$}%
             \mathrel{\mathop{\hbox to\wd\t@b@x{\rightarrowfill}}%
              \limits^{#1}}\,}
\def\leftarrowfill{$\m@th \mathord\leftarrow \mkern-6mu
     \cleaders\hbox{$\mkern-2mu \mathord- \mkern-2mu$}\hfill
      \mkern-6mu \mathord-$}
\def\froo#1{\setbox\t@b@x=\hbox{$\scriptstyle#1$}%
             \mathrel{\mathop{\hbox to\wd\t@b@x{\leftarrowfill}}%
              \limits^{#1}}\,}
 %
\def\frac#1#2{{#1\over#2}}
\def\frc#1#2{\relax\ifmmode{\textstyle{#1\over#2}} 
                    \else$#1\over#2$\fi}           
\def\inv#1{\frc{1}{#1}}                            
 %
\def\Claim#1#2#3{\bigskip\begingroup%
                  \xdef #1{\secsym\the\meqno}%
                   \writedef{#1\leftbracket#1}%
                    \global\advance\meqno by1\wrlabeL#1%
                     \noindent{\bf#2}\,#1{}\,:~\sl#3\vskip1mm\endgroup}

\def\QED{\rlx\hfill$\Box$\kern-7pt\raise3pt\hbox{$\surd$}\bigskip}
 %
 %
\def\1{\raise1pt\hbox{,}}     

\def\:{\buildrel!\over=}

\def\CP#1{\rlx\ifmmode\IP^{#1}\else\IP$^{#1}$\fi}
\def\cP#1{\rlx\ifmmode\IC{\rm P}^{#1}\else$\IC{\rm P}^{#1}$\fi}

\def\sll#1{\rlx\rlap{\,\raise1pt\hbox{/}}{#1}}
\def\Sll#1{\rlx\rlap{\,\kern.6pt\raise1pt\hbox{/}}{#1}\kern-.6pt}
%

 %
\def\eg{\hbox{\it e.g.}}        
 %
\def\topic#1{\bigskip\noindent$\2{\hbox{#1}}$\nobreak\vglue0pt%
              \noindent\ignorespaces}

\def\CY{Calabi-\kern-.2em Yau}

\def\3{\ifmmode\ldots\else$\ldots$\fi}
\def\\{\hfill\break}
\def\Z{\hfil\break\rlx\hbox{}\quad}
\def\3{\ifmmode\ldots\else$\ldots$\fi}

 %
 %

 %

 %

\def\NP#1{{\it Nucl.\,Phys.\,}{\bf#1\,}}
\def\PL#1{{\it Phys.\,Lett.\,}{\bf#1\,}}

\def\MPL#1{{\it Mod.\,Phys.\,Lett.\,}{\bf#1\,}}
\def\PRL#1{{\it Phys.\,Rev.\,Lett.\,}{\bf#1\,}}
\def\CMP#1{{\it Commun.\,Math.\,Phys.\,}{\bf#1\,}}

\def\IJMP#1{{\it Int.\,J.\,Mod.\,Phys.\,}{\bf#1\,}}
\baselineskip=13.0861pt plus2pt minus1pt
\parskip=\medskipamount
\let\ft=\foot
\noblackbox
 %
\def\Afour{\ifx\answ\bigans
            \hsize=16.5truecm\vsize=24.7truecm
             \else
              \hsize=24.7truecm\vsize=16.5truecm
               \fi}
 %
 %
\def\SaveTimber{\abovedisplayskip=1.5ex plus.3ex minus.5ex
                \belowdisplayskip=1.5ex plus.3ex minus.5ex
                \abovedisplayshortskip=.2ex plus.2ex minus.4ex
                \belowdisplayshortskip=1.5ex plus.2ex minus.4ex
                \baselineskip=12pt plus1pt minus.5pt
 \parskip=\smallskipamount
 \def\ft##1{\unskip\,\begingroup\footskip9pt plus1pt minus1pt\setbox%
             \strutbox=\hbox{\vrule height6pt depth4.5pt width0pt}%
              \global\advance\ftno by1
               \footnote{$^{\the\ftno)}$}{\ninepoint##1}%
                \endgroup}}
\catcode`@=12
%
%
%
 %
 %
\def\rd{{\rm d}}
\def\Ab{{\bar{A}}}

\def\\{\hfill\break}

\def\bm{{\ba\m}}

\def\bn{{\ba\n}}

\def\bs{{\ba\s}}

\def\rL{{\rm L}}
\def\rh{h}
 %
 %
\noblackbox
\pageno=0\nopagenumbers
\sequentialequations
\rightline{hep-th/9903114}
\vglue+35mm
 \centerline{\titlerm   Some Algebraic Symmetries}            \vskip4pt
 \centerline{\titlerm   of (2,2)-Supersymmetric Systems}      \vskip10pt
 \centerline{\titlerms  Tristan H\"ubsch\footnote{$^{\spadesuit}$}
       {On leave from the Institut Rudjer Bo\v{s}kovi\'c, Zagreb,
        Croatia.}}                                            \vskip-1pt
 \centerline{Department of Physics and Astronomy}             \vskip-3pt
 \centerline{Howard University, Washington, DC 20059}         \vskip-4pt
 \centerline{\tt thubsch\,@\,howard.edu}
\vfill

\centerline{ABSTRACT}\vskip2mm
\vbox{\leftskip=4.4em\rightskip=\leftskip\baselineskip=12pt\noindent
 The Hilbert spaces of supersymmetric systems admit symmetries which are
often related to the topology and geometry of the (target) field-space.
 Here, we study certain (2,2)-supersymmetric systems in 2-dimensional
spacetime which are closely related to superstring models. They all turn
out to posess some hitherto unexploidted and {\it geometrically and
topologically unobstructed\/} symmetries, providing new tools for studying
the topology and geometry of superstring target spacetimes, and so the
dynamics of the effective field theory in these.}
 \vskip+5mm
 \rightline{{\ninepoint\it TANSTAAFL?}}

\Date{Ides of March 1999\hfill}
\vfill\eject
\SaveTimber
 %
 %
\lref\rCGP{S.~Cecotti, L.~Girardello and A.~Pasquinucci:
       \NP{B328}(1989)701--722,\Z \IJMP{A6}(1991)2427.}

\lref\rDixon{L.~Dixon: in {\it Superstrings, Unified Theories and
       Cosmology 1987},\,p.67--127,\Z eds.~G.~Furlan et al.\ (World
       Scientific, Singapore, 1988).}

\lref\rTwJim{S.J.~Gates, Jr.: \PL{B352}(1995)43--49\semi
       S.J.~Gates, Jr., M.T.~Grisaru and M.E.~Wehlau: 
       \NP{B460}(1996)579--614.}

\lref\rGGRS{S.J.~Gates, Jr., M.T.~Grisaru, M.~Ro\v cek and
       W.~Siegel: {\it Superspace}\Z (Benjamin/Cummings Pub.\ Co.,
       Reading, Massachusetts, 1983).}

\lref\rCfld{P.S.~Green and T.~H\"ubsch: \PRL{61}(1988)1163--1167,\Z
       \CMP{119}(1989)431--441\semi
       P.~Candelas, P.S.~Green and T.~H\"ubsch:
       \PRL{62}(1989)1956--1959,\Z \NP{B330}(1990)49--102.}

\lref\rGP{B.R.~Greene and M.~R.~Plesser: \NP{B338}(1990)15-37.}

\lref\rGrHa{P.~Griffiths and J.~Harris: {\it Principles of Algebraic
       Geometry}\Z (John Wiley, New York, 1978).}

\lref\rMarjD{T.~H\"ubsch: \MPL{A6}(1991)1553--1559.}

\lref\rSing{T.~H\"ubsch: \MPL{A6}(1991)207--216.}

\lref\rHSS{T.~H\"ubsch: Haploid (2,2)-Superfields In 2-Dimensional
      Spacetime. hep-th/9901038.}

\lref\rSL{T.~H\"ubsch and S.-T.~Yau: \MPL{A7}(1992)3277.}

\lref\rSuSyM{E.~Witten: {\it J.\,Diff.\,Geom.\,\bf17}(1982)661--692.}

\lref\rPhases{E.~Witten, Nucl. Phys. B403 (1993) 159.}

 %
 %
\newsec{Introduction}\noindent
It has been known by now for quite some time~\rSuSyM\ that there exists a
formal but rather precise analogy between supersymmetry and exterior
calculus. This analogy derives from the fact that the generators of
supersymmetry are anticommuting and so nilpotent differential operators
just as exterior derivatives are.

We now turn to our case of interest: the 2-dimensional
(2,2)-supersymmetric gauged linear $\s$-models~\rPhases, exemplified
here by a simple Landau-Ginzburg/Calabi-Yau model. For simplicity, we
consider a simple hypersurface in a projective space; generalizations to
intersections of hypersurfaces in toric varieties and the corresponding
more general gauged linear $\s$-models are notationally tedious but
straightforward. Indeed, the same analysis will apply to gauged
$\s$-models with non-Abelian gauge symmetries, and so geometries of
complete intersections of hypersurfaces within non-abelian orbit spaces,
not just toric varieties. In the present case, the commuting canonical
coordinates are $\f^\m,\f^\bm$, with $\m=0,{\cdots},n$, the map immersing
the world sheet (Riemann surface) into the target space $\IP^n$ for which
the $\f^\m$ serve as homogeneous coordinates. The supersymmetric ground
states turn out to be further constrained to $X$, the hypersurface
$W(\f){=}0$ in $\IP^n$.

The anticommuting variables, $\j_\pm^\m,\j_\pm^\bm$ are local sections of
$K^{\pm1/2}{\8}\f^*(T_X)$ and $\bar{K}^{\pm1/2}{\8}\f^*({\ba T}_X)$,
where $K$ is the canonical bundle of the world-sheet. They satisfy the
equal-time anticommutation relations
\eqn\eETAR{\big\{\,\j_\pm^\m\,,\,\j_\pm^\bn\,\big\}={\rm g}^{\m\bn}~,}
where ${\rm g}^{\m\bn}$ is a metric on the target space $X$. Owing
to~\eETAR, half of the $\j_+$'s and half of the $\j_-$'s can be
interpreted as creation operators, the other half then being annihilation
operators. That is, there are two possible choices of Clifford-Dirac
vacua (and their conjugates):
\eqn\eCCV{ \j_+^\m\ket{0}_{cc}\id0\id\j_-^\m\ket{0}_{cc}~,\qquad
           \hbox{the $(c,c)$-vacuum}~, }
and
\eqn\eACV{ \j_+^\bm\ket{0}_{ac}\id0\id\j_-^\m\ket{0}_{ac}~,\qquad
           \hbox{the $(a,c)$-vacuum}~. }
Using these two distinct vacua, two distinct types of states (and their
conjugates) can be defined:
\eqn\eCCS{ \ket{\h;b,q}_{cc}=
            \h^{\m_1\cdots\m_b}_{\bn_1\cdots\bn_q}(\f,\bar\f\,)\>
             \j_{\m_1+}{\cdots}\j_{\m_b+}\,
              \j^{\bn_1}_-{\cdots}\j^{\bn_q}_-\ket{0}_{cc}~, }
where $\j_{\m+}\define {\rm g}_{\m\bn}\j_+^\bn$, and
\eqn\eACS{ \ket{\w;p,q}_{ac}=
            \w_{\m_1\cdots\m_p\bn_1\cdots\bn_q}(\f,\bar\f\,)\>
             \j^{\m_1}_+{\cdots}\j^{\m_p}_+\,
              \j^{\bn_1}_-{\cdots}\j^{\bn_q}_-\ket{0}_{ac}~. }

Upon the formal identification $\j_-^\bm\sim\rd z^\bm$,
$\j_{\m+}\sim\vd_\m$ and $\j_+^\m\sim\rd z^\m$, we have that
$\ket{\h;b,q}$ correspond to $\wedge^bT_X$-valued Dolbeault $q$-forms,
whereas $\ket{\w;p,q}$ correspond to $(p,q)$-forms. Furthermore, those
$\ket{\h;b,q}$ and $\ket{\w;p,q}$ which are annihilated by all
supercharges represent zero-energy states, correspond to
{\it harmonic\/} forms, and so encode information about global geometry
and topology of the hypersurface $W(\f){=}0$ within $\IP^n$.

\newsec{Two Ubiquitous $SL(2,\IC)$ Actions}\noindent
The case at hand falls in a very well studied category: the hypersurface
$W(\f){=}0$ in $\IP^n$ is K\"ahler. Now, the $p,q$-forms on all K\"ahler
manifolds admit a so-called Lefschetz $SL(2)$ action, depending, besides
the canonical wedge product and Hodge star operator, only on the choice of
the K\"ahler metric~\rGrHa. The fermionic analogue of this $SL(2)$ action
has been known for some time (see Refs.~\refs{\rCGP,\rSL}, and references
therein).

\subsec{The standard Lefschetz $SL(2,\IC)$ algebra}\noindent
Acting on the wave-functions~\eCCS, \eACS\ and their conjugates, we
define two ``ladder'' operators:
\eqna\eVL
 $$
    \rL_+ \define {\rm g}_{\m\bn}\j_+^\m\j_-^\bn~, \qquad
    \rL_- \define {\rm g}_{\m\bn}\j_-^\m\j_+^\bn~, \eqno\eVL{a,b}
 $$
and calculate their commutator:
 $$ \eqalignno{ \rh
 &\define\big[\, \rL_+ \,,\, \rL_- \,\big]
  = {\rm g}_{\m\bn}{\rm g}_{\r\bs}
      \big[\j_+^\m\j_-^\bn\,,\,\j_-^\r\j_+^\bs\big]~,\cr
 &={\rm g}_{\m\bn}{\rm g}_{\r\bs}
      \Big(\j_+^\m\big\{\j_-^\bn\,,\,\j_-^\r\big\}\j_+^\bs
    -\j_-^\r\big\{\j_+^\m\,,\,\j_+^\bs\big\}\j_-^\bn\Big)~,\cr
 &={\rm g}_{\m\bn}
      \big(\j_+^\m\j_+^\bn - \j_-^\m\j_-^\bn\big)~. &\eVL{c}\cr}
 $$
Using the anticommutation relations~\eETAR, this expression can be
brought into the more standard form:
 $$
   \rh={\rm g}_{\m\bn}\big(\j_+^\m\j_+^\bn +
                            \j_-^\bn\j_-^\m\big)-(n{+}1)~, \eqno\eVL{d}
 $$
where now the creation operators (when acting on
$\oplus_{p,q}\ket{\w;p,q}_{ac}$) are to the left of the annihilation
operators. The monomials $\j_+^\m\j_+^\bn$ and $\j_-^\bn\j_-^\m$ are
simply (fermion) number operators, and $\rh$ then simply stands for the
total (fermion) number operator, shifted so that its eigenvalues on
$\ket{\w;p,q}$ range from $(p{+}q)=-(n{+}1)$ to $(p{+}q)=(n{+}1)$.

A similar calculation verifies that
\eqn\eXXX{ \big[\, \rh \,,\, \rL_\pm \,\big] = \pm2\rL_\pm~, }
whence $\{\rL_\pm,\rh\}$ form an $SL(2)$ algebra. It is similarly
easy to verify that:
\eqn\eXXX{ \rL_\pm\ket{\w;p,q}_{ac} \mapsto
            \ket{\w';p{\pm}1,q{\pm}1}_{ac}~, }
so that this $SL(2)$ action coincides with the standard Lefschetz $SL(2)$
action. In the usual layout of the Hodge diamond, where the
$(p,q)=(0,0)$- and $(n,n)$-forms are at the bottom and top corners,
respectively, this $SL(2)$ acts {\it vertically\/}. In fact,
$\j_+^\m,\j_-^\bn$ play the r\^oles of Griffiths and Harris's {\it
formal\/} basis elements $e_k,\bar e_k$, while $\j_-^\m,\j_+^\bn$ play the
roles of their {\it duals\/}, $i_k,\bar i_k$~\rGrHa. Finally, we
complexify $g_{\m\bn}\to(g_{\m\bn}{+}iB_{\m\bn})$, using the
antihermitian 2-form $B$ familiar from 2-dimensional $\s$-models related
to string theory. The $SL(2)$ transformation parameters thus become
complex, producing the {\it complexified\/} Lefschetz $SL(2,\IC)$ action.

\subsec{The mirror $SL(2,\IC)$ algebra}\noindent
The existence of the mirror map among (families of) Calabi-Yau 3-folds
implies that there exists a $Y$, the mirror model of the manifold $X$,
such that $H^q(X,\wedge^pT^*_X)\approx H^q(Y,\wedge^pT_Y)$ and
$H^q(X,\wedge^pT_X)\approx H^q(Y,\wedge^pT^*_Y)$; note that
$\wedge^pT_X=\wedge^{n-p}T^*_X$ on Calabi-Yau $n$-folds. Therefore, the
mirror map identifications may also be stated as  $H^{p,q}(X)\approx
H^{n-p,q}(Y)$. The standard Lefschetz $SL(2)$ action on $H^{p,q}(X)$ is
then mapped to an action on $H^{n-p,q}(Y)$, where it now acts
{\it horizontally\/}! Similarly, the standard Lefschetz $SL(2)$ action on
$H^{p,q}(Y)$ has a pre-image on $H^{n-p,q}(X)$, where it acts
horizontally.

In the Landau-Ginzburg model for $X$, this horizontal Lefschetz-like
$SL(2)$ action is easy to identify. Recall that $X$ is defined as the
$W{=}0$ hypersurface within $\IP^n$; let then $W_{\m\n}=\vd_\m\vd_\n W$.
We now define another two `ladder' operators:
\eqna\eHL
 $$
    \G_+ \define W_{\bm\,\bn}\j_+^\bm\j_-^\bn~,\quad
    \G_- \define W_{\m\n}\j_-^\m\j_+^\n~, \eqno\eHL{a,b}
 $$
where $W_{\bm\,\bn}\define {\rm g}_{\k\bm}W^{\k\s}{\rm g}_{\s\bn}$, with
$W^{\k\s}$ being the matrix-inverse of $W_{\k\s}$:
$W_{\m\n}W^{\m\s}=\d^\s_\n$. Furthermore,
 $$
    \m \define \big[\, \G_+ \,,\, \G_- \,\big]
    = {\rm g}_{\m\bn}\big[\j_+^\bn\j_+^\m-\j_-^\m\j_-^\bn\big]~. 
\eqno\eHL{c}
 $$
Again, we may rewrite this as in the more standard way as
 $$
    \m = {\rm g}_{\m\bn}\big[\j_+^\bn\j_+^\m+\j_-^\bn\j_-^\m\big]
         -(n{+}1)~.  \eqno\eHL{d}
 $$
Again, the creation operators (now when acting on
$\oplus_{b,q}\ket{h;b,q}_{cc}$) are to the left of the annihilation
operators. The monomials $\j_+^\bn\j_+^\m$ and $\j_-^\bn\j_-^\m$ are again
the (fermion) number operators, and $\m$ then simply stands for the total
(fermion) number operator, shifted so that its eigenvalues on
$\ket{h;b,q}$ range from $(b{+}q)=-(n{+}1)$ to $(b{+}q)=(n{+}1)$.

A quick calculation verifies that
\eqn\eXXX{ \big[\, \m \,,\, \G_\pm \,\big] = \pm2\G_\pm~, }
whence $\{\G_\pm,\m\}$ form another $SL(2)$ algebra. It is similarly
easy to verify that:
\eqn\eXXX{ \G_\pm\ket{\w;p,q}_{ac} \mapsto
                  \ket{\w';p{\mp}1,q{\pm}1}_{ac}~, }
so that this second $SL(2)$ action coincides with the mirror map pre-image
of the standard Lefschetz $SL(2)$ action on the mirror, $Y$; it acts
{\it horizontally} on the Hodge diamond of $X$.

However, note that the action of the ladder operators $\rL_\pm$ and
$\G_\pm$ is swapped when acting on the $\ket{\h;b,q}_{cc}$:
\eqn\eXXX{{\eqalign{
 \rL_\pm\ket{\h;b,q}_{cc} &\mapsto \ket{\h';b{\mp}1,q{\pm}1}_{cc}~,
 \qquad\hbox{horizontal action};\cr
 \G_\pm\ket{\h;b,q}_{cc} &\mapsto \ket{\h';b{\pm}1,q{\pm}1}_{cc}~,
 \qquad\hbox{vertical action}.\cr
 }}}

Moreover, straightforward calculations show that these two $SL(2)$ actions
commute, whence $\{\rL_\pm,\rh\}$ and $\{\G_\pm,\m\}$ generate an
$SL(2)_L{\times}SL(2)_\G$. On $\oplus_{p,q}\ket{\w;p,q}_{ac}$, the first
factor acts vertically and the second one horizontally, while on
$\oplus_{b,q}\ket{\h;b,q}_{cc}$ their actions are swapped. Therefore,
$SL(2)_\G$ generated by $\{\G_\pm,\m\}$ is the (mirror map pre-image of
the) mirror of $SL(2)_L$ generated by $\{\rL_\pm,\rh\}$.

\vbox{
$$\vbox{\offinterlineskip
\hrule height 1.1pt
\halign{&\vrule width 1.1pt#&\strut\quad\hfil#\hfil\quad&
\vrule#&\strut\quad\hfil#\hfil\quad&
\vrule#&\strut\quad\hfil#\hfil\quad&\vrule width 1.1pt#\cr
height3pt&\omit&&\omit&&\omit&\cr
 &{\bf Generators}&&{\bf on} $\ket{\w;p,q}_{a,c}$&
                   &{\bf on} $\ket{\w;b,q}_{c,c}$&\cr
height3pt&\omit&&\omit&&\omit&\cr
\noalign{\hrule height 0.8pt}
height3pt&\omit&&\omit&&\omit&\cr
 &$\big\{\,\rL_+\,,\,\rL_-\,,\,h\,\big\}$&
 & Lefschetz $SL(2,\IC)$ && `mirror' $SL(2,\IC)$ &\cr
height3pt&\omit&&\omit&&\omit&\cr
\noalign{\hrule}
height3pt&\omit&&\omit&&\omit&\cr
 &$\big\{\,\G_+\,,\,\G_-\,,\,\m\,\big\}$&
 & `mirror' $SL(2,\IC)$ && Lefschetz $SL(2,\IC)$ &\cr
height3pt&\omit&&\omit&&\omit&\cr} \hrule height
1.1pt}$$ \vskip7pt}

Finally, it is obvious that the Bogoliubov transformation
$\j_+^\m\iff\j_+^\bm$ becomes exactly the `relative sign change' in the
action of the $U(1)_L{\times}U(1)_R$ of the corresponding superconformal
field theory, and so {\it is} the mirror
map~\refs{\rDixon,\rGP;\rSL,\rTwJim}. It is equally clear that the
same field redefinition also swaps the two $SL(2)$ actions,
$\{\rL_+,\rL_-,h\}\iff\{\G_+,\G_-,\m\}$, proving that these are indeed
{\it the} mirror (pre)images of each other; see also Ref.~\refs{\rHSS}.

\newsec{Discussion}\noindent
The main result proven, we now address some additional issues in turn.

\subsec{The mirror map and marginal operators}\noindent
The definition of $\{\G_\pm,\m\}$ uses, most crucially, the Hessian of the
defining polynomial, $W$. Notice that ${\rm g}_{\m\bn}$ is in fact a
K\"ahler metric, and so also a Hessian: ${\rm g}_{\m\bn}=\vd_\m\vd_\bn K$.
Since the two $SL(2,\IC)$ algebras are mirror (pre)images of each other,
the K\"ahler potential $K$ for the metric ${\rm g}_{\m\bn}$ and the
defining polynomial (superpotential) $W$ must be mirrors of each other.
 In 2-dimensional (2,2)-superspace, the `K\"ahler potential' function $K$
is defined only up to the addition of terms each of which is annihilated at
least by one of the four superderivatives~\refs{\rHSS}. This
`undefinedness' is far larger than in spacetimes of more than 2
dimensions! Also unlike its familiar 4-dimensional counterpart, the
superpotential in 2-dimensions is similarly `undefined', although in more
restrictive way~\refs{\rHSS}.

Furthermore, the definition of $\G_-$ involves the matrix {\it inverse\/}
of the Hessian of $W$. This exists provided the determinant of the Hessian
is non-zero, and which allows $W$ to be mildly singular: $\rd W$ may
vanish, as long as the locus of $\rd W{=}0$ are only nodes (double points). 
Mirror symmetry then implies that $K$ may be `singular' in the sense that
$\rd K$ may vanish, as long as the (Hermitian) matrix of second
derivatives, ${\rm g}_{\m\bn}$, remains invertible. But this, and nothing
more is precisely the `standard' requirement of the K\"ahler potential!
 So, since one never expects anything more of $K$, mirror symmetry suggests
that:
\nobreak\vglue1mm\vbox{\narrower
 \item{$\bullet$} Superpotentials should also be allowed to singularize,\\
as long as their Hessians are invertible~\refs{\rSing,\rCfld}.}\bigskip

In the (2,2)-supersymmetric field theory, the mirror relation between the
K\"ahler potential and the superpotential may come as a surprise.
While the latter is a purely chiral function, the former is a neither
chiral nor anti-chiral, but real. Whereas the latter enters the Lagrangian
as an $F$-term and does not renormalize~\ft{See, \eg, p.358 of
Ref.~\refs{\rGGRS}\ for an important caveat to this `theorem'.}, the
former figures in a $D$-term, not protected by the usual
non-renormalization theorems. However, this real function does give rise
to a collection of {\it twisted\/}-chiral marginal operators (one for
each $(a,c)$-modulus), just as the superpotential produces a collection
of chiral marginal operators (one for each $(c,c)$-modulus)~\rMarjD. Of
course, the Bogoliubov transformation $\j_+^\m\iff\j_+^\bm$ (a.k.a.\
mirror map) also swaps the chiral and the twisted-chiral fields, again
verifying that:
\nobreak\vglue1mm\vbox{\narrower
 \item{$\bullet$}$D$-terms can yield {\it twisted}-chiral marginal
operators,\\ the mirror map (pre)images of the $F$-term chiral marginal
operators.}\bigskip

Quite importantly, the definition of the $SL(2,\IC)_L{\times}SL(2,\IC)_\G$
algebra is purely algebraic. Thus, it `comes for free', in {\it any\/}
(2,2)-supersymmetric model that features a metric $g_{\m\bn}$ and a
superpotential $W$. Geometrically, the $SL(2,\IC)_L{\times}SL(2,\IC)_\G$
is {\it unobstructed\/} since it acts on the  contractible fibres of
the bundle $\wedge(T{\times}T^*)_X$.
 In gauged models, $g_{\m\bn}$ is defined upon passing to a `gauge slice':
e.g., in Witten's gauged linear $\s$-model, the gauging of the various
$U(1)$ symmetries induces the generalization of the Fubini-Study metric on
the gauge quotient toric variety within which the hypersurface $W(\f){=}0$
lies.

Also, note that the $SL(2,\IC)_L{\times}SL(2,\IC)_\G$ algebra
is a (small) part of what are generally known as `dynamical
symmetry'/`spectrum-generating' algebras. That is,
\nobreak\vglue1mm\vbox{\narrower
 \item{$\bullet$}Given a (judiciously chosen) quarter of the
supersymmetric $\ket{\w;p,q}_{ac}$'s, the others are obtained by
applying the $SL(2,\IC)_L{\times}SL(2,\IC)_\G$ ladder operators.}

\subsec{Extensions}\noindent
The $SL(2,\IC)_L{\times}SL(2,\IC)_\G$ symmetry found above may be extended
in several ways.

\topic{More fermions}\noindent
Clearly, $N$-fold extended (2,2)-supersymmetry will give rise to $N$
`species' of fermions, each of which having $n$ fermions. A more
complicated situation occurs in the more general models of
Ref.~\refs{\rHSS}, where the different species of fermions stem from
different superfields, and are therefore not required to be equal in
number. Returning to the simple case of $N$-extended supersymmetry,
operators of the type~\eVL{} and~\eHL{} now become $N{\times}N$ matrix
operators in the `species space':
\eqna\eVHL
 $$\cmathno{
    \rL^i_{j+} \define {\rm g}_{\m\bn}\j_+^{i\m}\j_{j-}^\bn~, \quad
    \rL^i_{j-} \define {\rm g}_{\m\bn}\j_-^{i\m}\j_{j+}^\bn~,
                                                            &\eVHL{a,b}\cr
    \rh^i_j \define {\rm g}_{\m\bn}
     \big[\j_+^{i\m}\j_{j+}^\bn - \j_-^{i\m}\j_{j-}^\bn\big]~,
                                                            &\eVHL{c}\cr
 }$$
and Eqs.~\eHL{} now become
 $$\cmathno{
    \G_{ij+} \define W_{\bm\,\bn}\j_{i+}^\bm\j_{j-}^\bn~,\quad
    \G_-^{ij} \define W_{\m\n}\j_-^{i\m}\j_+^{j\n}~,        &\eVHL{d,e}\cr
    \m^i_j \define {\rm g}_{\m\bn}
     \big[\j_{j+}^\bn\j_+^{i\m}-\j_-^{i\m}\j_{j-}^\bn\big]~.
                                                            &\eVHL{f}\cr
 }$$
We will also need:
 $$ \twoeqsalignno{
 H^{ij}&\define W_{\m\r} \j_+^{i\m}\j_+^{j\r}~,
 \quad&\quad
 H_{ij}&\define W_{\bn\,\bs} \j_{i+}^\bn\j_{j+}^\bs~, &\eVHL{g,h}\cr
 I^{ij}&\define W_{\m\r} \j_-^{i\m}\j_-^{j\r}~,
 \quad&\quad
 I_{ij}&\define W_{\bn\,\bs} \j_{i-}^\bn\j_{j-}^\bs~, &\eVHL{i,j}\cr
 }$$
which are antisymmetric in $i,j$ and so vanish when there is a single
species of fermions. Also, it will be convenient to use
 $$ \twoeqsalignno{
 J^i_j&\define {\rm g}_{\m\bn}\j^{i\m}_+\j_{j+}^\bn~, \quad\hbox{and}\quad
 \quad&\quad
 J^i_j&\define {\rm g}_{\m\bn}\j^{i\m}_-\j_{j-}^\bn~, &\eVHL{k,l}\cr
 }$$
so that
\eqn\eXXX{ \rh^i_j = J^i_j - K^i_j~,\quad\hbox{and}\quad
           \m^i_j = (n{+}1)\d^i_j - J^i_j - K^i_j~. }

We now find that the algebra spanned by
$\{\Ione,\rL_\pm,\G_\pm,J,K,H^{..},H_{..},I^{..},I_{..}\}$ is:
\eqna\eClc{\ninepoint
 $$ \twoeqsalignno{
 \big[ \rL^i_{j+} , \rL^k_{l-} \big]
  &=~\d^k_j J^i_l - \d^i_l K^k_j~, &
 \big[ \rL^i_{j+} , \G_{kl+} \big] &=~\d^i_l I_{kj}~, &\eClc{a}\cr
 \big[ \rL^i_{j+} , \G^{kl}_- \big] &=~\d^k_j H^{il}~, &
 \big[ \rL^i_{j-} , \G_{kl+} \big] &=~H_{jk} \d^i_l~, &\eClc{b}\cr
 \big[ \rL^i_{j-} , \G^{kl}_- \big] &=~ -I^{ik} \d^l_j~, &
 \big[ \G_{jl+} , \G^{ik}_- \big]
 &=~\d^k_j(\6(n{+}1)\d^i_l{-}J^i_l) - \d^i_l K^k_j~,          &\eClc{c}\cr
 \big[ H^{ij} , H_{kl} \big]
 &=~-\d^{[i}_{[k}J^{j]}_{l]} + (n{+}1)\d^{[ij]}_{kl}\Ione~, &
 \big[ I^{ij} , I_{kl} \big]
 &=~-\d^{[i}_{[k}K^{j]}_{l]} + (n{+}1)\d^{[ij]}_{kl}\Ione~,   &\eClc{d}\cr
 \big[ H_{ij} , \rL^k_{l+} \big]
 &=~\d^k_j\G_{il+} - \d^k_i\G_{jl+}~, &
 \big[ I^{ij} , \rL^k_{l+} \big]
 &=~\d^i_l\G^{jk}_- - \d^j_l\G^{ik}_-~,                       &\eClc{e}\cr
 \big[ H^{ij} , \rL^k_{l-} \big]
 &=~\d^j_l\G^{ki}_- - \d^i_l\G^{kj}_-~, &
 \big[ I_{ij} , \rL^k_{l-} \big]
 &=~\d^k_j\G_{li+} - \d^k_i\G_{lj+}~,                         &\eClc{f}\cr
 \big[ H^{ij} , \G_{kl+} \big]
 &=~\d^j_k\rL^i_{l+} - \d^i_k\rL^j_{l+}~, &
 \big[ I^{ij} , \G_{kl+} \big]
 &=~\d^i_l\rL^j_{k-} - \d^j_l\rL^i_{k-}~,                     &\eClc{g}\cr
 \big[ H_{ij} , \G^{kl}_- \big]
 &=~\d^l_j\rL^k_{i-} - \d^l_i\rL^k_{j-}~, &
 \big[ I_{ij} , \G^{kl}_- \big]
 &=~\d^k_i\rL^l_{j+} - \d^k_j\rL^l_{i+}~,                     &\eClc{h}\cr
 \big[ J^i_j , \rL^k_{l+} \big] &=~\d^k_j\rL^i_{l+}~, &
 \big[ J^i_j , \rL^k_{l-} \big] &=~-\d^i_l\rL^k_{j-}~,        &\eClc{i}\cr
 \big[ J^i_j , \G_{kl+} \big] &=~\d^i_k\G_{jl+}~, &
 \big[ J^i_j , \G^{kl}_- \big] &=~-\d^l_j\G^{ki}_-~,          &\eClc{j}\cr
 \big[ K^i_j , \rL^k_{l+} \big] &=~-\d^i_l\rL^k_{j+}~, &
 \big[ K^i_j , \rL^k_{l-} \big] &=~\d^k_j\rL^i_{l-}~,         &\eClc{k}\cr
 \big[ K^i_j , \G_{kl+} \big] &=~-\d^i_l\G_{kj+}~, &
 \big[ K^i_j , \G^{kl}_- \big] &=~\d^k_j\G^{il}_-~,           &\eClc{l}\cr
 \big[ J^i_j , H^{kl} \big] &=~H^{i[l}\d^{k]}_j~, &
 \big[ J^i_j , H_{kl} \big] &=~\d^i_{[k}H_{l]j}~,             &\eClc{m}\cr
 \big[ K^i_j , I^{kl} \big] &=~I^{i[l}\d^{k]}_j~, &
 \big[ K^i_j , I_{kl} \big] &=~\d^i_{[k}I_{l]j}~,             &\eClc{n}\cr
 \big[ J^i_j , J^k_l \big] &=~\d^k_j J^i_l - \d^i_l J^k_j~, &
 \big[ K^i_j , K^k_l \big] &=~\d^k_j K^i_l - \d^i_l K^k_j~,   &\eClc{o}\cr
 }$$}%
all other commutators being zero. Note that the identity, $\Ione$, appears
on the right-hand sides of both Eqs.~\eClc{d}, and so must be included as
a generator of the algebra; it of course commutes with all other
generators.

\topic{Superalgebras}\noindent
Besides the bosonic operators bilinear in the fermions~\eVL{} and~\eHL{},
we can introduce fermionic operators, linear in the $\j,\bar\j$:
\eqn\eVF{ {\cmath{ 
 A_+ = A_\m\j^\m_+~,\qquad \Ab_+ = A_\bn\j^\bn_+~, \cr
 A_- = A_\m\j^\m_-~,\qquad \Ab_- = A_\bn\j^\bn_-~, \cr
 }}}
Their anticommutators must be expressible as a linear combination of
$\Ione,L_\pm,\rh,\G_\pm,\m$, and the vector space $A,\Ab,B,\bar{B}$
must form a representation of $SL(2,\IC)_L{\times}SL(2,\IC)_\G$. The latter
requirement forces the commutators of the operators~\eVL{} with the~\eVF\
to be expanded over the~\eVF. The first requirement produces
\eqn\eAAb{ \{A_+,\Ab_+\} = A_\m A_\bn g^{\m\bn}\Ione~\id~\|A\|^2\Ione~=~
           \{A_-,\Ab_-\}~. }
Next, we find
\eqn\eSLL{{\cmath{
[L_+,A_+]=0~,\quad [L_-,A_+]=+A_-~,\quad [\rh,A_+]=+A_+~,\cr
[L_+,A_-]=+A_+~,\quad [L_-,A_-]=0~,\quad [\rh,A_-]=-A_-~,\cr
[L_+,\Ab_+]=-\Ab_-~,\quad [L_-,\Ab_+]=0~,\quad [\rh,\Ab_+]=-\Ab_+~,\cr
[L_+,\Ab_-]=0~,\quad [L_-,\Ab_-]=-\Ab_+~,\quad [\rh,\Ab_-]=+\Ab_-~;\cr
 }}}
and
\eqn\eSLG{{\cmath{
[\G_+,A_+]=-\Ab'_-~,\quad [\G_-,A_+]=0~,\quad [\m,A_+]=-A_+~,\cr
[\G_+,A_-]=+\Ab'_+~,\quad [\G_-,A_-]=0~,\quad [\m,A_-]=-A_-~,\cr
[\G_+,\Ab_+]=0~,\quad [\G_-,\Ab_+]=+A'_-~,\quad [\m,\Ab_+]=+\Ab_+~,\cr
[\G_+,\Ab_-]=0~,\quad [\G_-,\Ab_-]=-A'_+~,\quad [\m,\Ab_-]=+\Ab_-~,\cr
 }}}
with
\eqn\eTrs{
  A'_\pm\define(W_\m{}^\bn A_\bn)\j^\m_\pm~, \quad\hbox{and}\quad
  \Ab'_\pm\define(W_\bn{}^\m A_\m)\j^\bn_\pm~. }
The matrices $W_\m{}^\bn\define W_{\m\r}{\rm g}^{\r\bn}$ and
$W_\bn{}^\m\define W_{\bn\,\bs}{\rm g}^{\m\bs}$ act as a (conjugating)
linear transformation on the coefficient functions
$A_\m,A_\bn,\Ab_\m,\Ab_\bn$; they are well-defined since the Hessians
$W_{\m\r}$ and ${\rm g}_{\m\bn}$ are invertible. The relations~\eSLL\
assure that $A_\pm$ and $\Ab_\pm$ transform as two $SL(2,\IC)_L$
spin-$\inv2$ doublets, and~\eSLG\ show that $(A_+,\Ab_-)$ and
$(A_-,\Ab_+)$ are $SL(2,\IC)_\G$ doublets, twisted by the
$W_.^.$-transformation.
 This guarantees that $\{A_\pm,\Ab_\pm;\Ione,L_\pm,\rh,\G_\pm,\m\}$
generate a supergroup.

Of course, it is also possible to expand the right-hand side of the
anticommutators~\eAAb\ over non-trivial (differential) operators over the
bosonic degrees of freedom. One such possibility leads to the
well-studied field-space {\it supersymmetry\/} algebra; see for example
Ref.~\refs{\rCGP}. If the target manifold $X$ admits null-vectors, the
$A_\m,A_\bn$ may be chosen to have zero norm, whereupon they generate a
BRST-like subalgebra of the superalgebra discussed in Eqs.~\eAAb--\eTrs.
Another possibility is to let the $A_\m,A_\bn$ take values in a
non-Abelian Lie algebra. These and other such extensions are left for
another occasion.

 %
\vfill
\bigskip\noindent{\it Acknowledgments\/}:
I am indebted to the Department of Energy for their generous support
through the grant DE-FG02-94ER-40854. Many thanks also to the Institute
for Theoretical Physics at Santa Barbara, where part of this work was done
with the support from the National Science Foundation, under the Grant
No.~PHY94-07194.

\vfill
\listrefs

\bye